\documentclass[%
 reprint,
 amsmath,amssymb,
 aps,
]{revtex4-2}

\usepackage{graphicx}
\usepackage{dcolumn}
\usepackage{bm}

\begin{document}

\title{Fast, parallel, three-dimensional rearrangement of atom arrays using superpositions of elementary holograms}

\author{D. J. Hoare}
\author{P. Dunniece}
\author{J. Rodewald}
\author{K. K. Voges}
\altaffiliation[Current address: ]{Institut für Quantenoptik, Leibniz Universität Hannover, 30167 Hannover, Germany}
\author{M. R. Tarbutt}
\affiliation{Centre for Cold Matter, Blackett Laboratory,  Imperial College London, Prince Consort Road, London SW7 2AZ, UK}

\begin{abstract}
 Assembling defect-free arrays of neutral atoms requires real-time rearrangement of stochastically loaded optical tweezers. Conventional rearrangement methods shuttle atoms sequentially, whereas high-speed spatial light modulators (SLMs) can move all atoms in parallel, which scales more favorably as the array size grows.  For fast parallel rearrangement, the sequence of holograms displayed on the SLM should be computed more rapidly than the SLM refresh rate, and there should be phase continuity between successive holograms to minimize transient intensity fluctuations. We calculate holograms analytically by adding together the complex fields needed to produce each individual trap, and we ramp the phase of each trap linearly as it moves to its target location. 
 Trap positions enter as continuous variables, rather than tied to points on a grid, so the method provides high positioning accuracy.
 Using a single GPU, we compute each intermediate hologram in $0.284\,$ms, considerably faster than the SLM refresh time of $0.696\,$ms, so the rearrangement speed is hardware-limited. In each step, we use the free processing time to iteratively optimize the holograms for equal intensity across the traps, decreasing trap non-uniformity to $1.24\%$ during rearrangement at no cost in total time. Starting from an array of 256 sites loaded stochastically with $^{87}$Rb atoms, we assemble a 100-atom array in under 20~ms with a rearrangement survival probability of $99.31(12)\%$ per atom. This rearrangement increases the temperature of the atoms by $9(3)\,\mu$K, which is small compared to the trap depth of $480~\mu$K. Our method can also move individual atoms axially, enabling transport in different planes. We demonstrate this capability with a rearrangement protocol where one set of atoms leapfrogs over another. 
\end{abstract}

\maketitle

\section{Introduction}
\label{sec:intro}

In recent years, optical tweezer arrays of neutral atoms have emerged as a leading platform for quantum computing and simulation. The atoms provide second-scale coherence times~\cite{manetsch2025tweezer}, and fast, high-fidelity quantum logic operations~\cite{evered2023high, ma2023high} through excitation to Rydberg states. A particular strength of this architecture lies in its dynamic reconfigurability~\cite{Barredo2016, barredo2018synthetic, Lu2026a} and scalability, with the maximum number of physical qubits bounded primarily by available laser power. Holographic tweezer arrays of thousands of atomic qubits have been demonstrated~\cite{manetsch2025tweezer, Lin2025}, and meta-surface tweezer arrays that trap more than 10,000 atoms have been developed~\cite{wang2026trapping}. In quantum simulation, recent highlights include simulations of quantum magnetism, phase transitions, and topological spin liquids~\cite{semeghini2021probing, ebadi2021quantum, Scholl2021}. In neutral atom quantum computing, recent achievements include two-qubit entangling gate fidelities of 99.5\% applied across tens of atoms in parallel, surpassing the surface-code threshold for error correction~\cite{evered2023high}, and advanced error-correction protocols such as real-time mid-circuit readout and erasure conversion~\cite{ma2023high}. Combining these capabilities with the all-to-all connectivity enabled by the coherent transport of atomic qubits~\cite{beugnon2007two, bluvstein2022quantum, bluvstein2024logical} firmly establishes neutral atom tweezer arrays as a promising architecture for scalable, fault-tolerant quantum computation. 

The process of loading single atoms into tweezers is inherently stochastic, typically yielding trap occupancies of 50-60\%, due to light-assisted collisions between pairs of atoms~\cite{Schlosser2001, Schlosser2002}. This initial array must then be rearranged to produce a defect-free array. Similarly, implementing advanced quantum computing paradigms that use distinct computational zones~\cite{bluvstein2022quantum} and all-to-all connectivity requires highly efficient, massively parallel mid-circuit atom transport. Traditionally, the assembly of defect-free arrays has been achieved using moveable tweezers generated by acousto-optic deflectors (AODs) to shuttle atoms sequentially~\cite{Barredo2016}. This approach incurs a time overhead that scales linearly with the number of atoms. By contrast, high-speed spatial light modulators (SLMs) provide a route to constant-time scaling by enabling the parallel movement of all atoms simultaneously, provided the computational overhead of calculating these underlying holographic phase masks in real time can be overcome.

Computing the hologram that generates a desired tweezer array is a phase-retrieval problem, typically solved using an iterative Fourier transform (IFT) algorithm, such as the weighted Gerchberg-Saxton (WGS) algorithm \cite{fienup1982phase, di2007computer}. These algorithms can generate highly uniform arrays, but they leave the optical phase of each tweezer unconstrained, treating them as free parameters that evolve randomly as the algorithm converges. This is not an issue for stationary arrays, but poses a problem for generating intermediate holograms in a rearrangement sequence. If, for a given tweezer, the phase difference between consecutive holograms is too large, the transient optical fields during the SLM refresh will interfere destructively, causing fluctuations in trap depth that result in atom loss. 
Knottnerus \textit{et al.}~\cite{Knottnerus2025} address this by explicitly controlling the optical phase of the tweezers and using linear phase interpolation (LPI) to minimize the phase difference between successive holograms. In their approach, initial and final holograms are optimized using the WGS algorithm, and trap phases are extracted by a fast Fourier transform (FFT). Intermediate holograms are then calculated by linearly interpolating the tweezer positions and phases and applying a single inverse FFT at each step. This method achieved a rearrangement survival probability of 99.6\% per atom with a computation-to-display time of 2.736~ms per intermediate hologram \cite{Knottnerus2025}. However, because the FFT samples the focal plane on a discrete grid, the positional resolution is constrained to a Fourier unit. Higher resolution can be achieved using zero-padding to expand the computational space and reduce the Fourier unit, but at the expense of increasing computation time \cite{Kim2019, Knottnerus2025}. Lin \textit{et al.}~\cite{Lin2025} show that this overhead can be reduced by using a convolutional neural network (CNN) and two parallel GPUs to generate holograms with positional accuracy equivalent to a 64-times expanded computational space. Using this method, they rearrange atoms in 60~ms and show that this time remains fixed as the array size increases. 

In this work, we combine LPI with a superposition-of-elementary-holograms (SEH) approach to hologram computation, in which holograms are calculated analytically. The SEH method is well established in holographic optical trapping \cite{liesener2000multifunctional, di2007computer, chew2024ultraprecise}; what is specific to this work is its implementation in a real-time rearrangement protocol. In this approach, a hologram is calculated by summing the complex field contributions of each tweezer and taking the argument of the result. The complex field for each tweezer is defined by a spatial phase gradient with a wavevector that sets its exact position in the Fourier plane, along with a phase offset and amplitude weight. Because each wavevector is a continuous parameter, tweezer positions are not fixed to a discrete computational grid, so the SEH approach provides precise positional accuracy~\cite{chew2024ultraprecise}.

Using a single commercially available GPU (NVIDIA RTX 5090), we compute each hologram in 0.284~ms, considerably faster than the 0.696~ms physical refresh cycle of the SLM. As such, our rearrangement speed is hardware-limited rather than limited by computation. 
Consequently, there is enough time to complete an additional intensity-equalization iteration per rearrangement step, reducing trap intensity non-uniformity during rearrangement from $20\%$ to $1.24\%$ while remaining hardware-limited. We numerically verify that phase continuity is satisfied, finding a standard deviation in phase error of $0.0432$~rad with no intensity equalization iterations per step and $0.045$~rad with two, an order-of-magnitude decrease from the $0.2~\rm rad$ reported in Ref.~\cite{Lin2025}. We assemble 100-atom arrays from stochastically loaded 256-site arrays, measuring a rearrangement survival probability of $99.31(12)\%$ with a total rearrangement time under 20 ms. We measure a rearrangement-induced temperature increase of $9(3)\,\mu$K, which, with a trap depth of $480\,\mu\rm K$, shows that the transported atoms remain cold and deeply confined. We extend the transport into three dimensions (3D), dividing lattices into sub-lattices that are transported into different axial planes, and measuring a round-trip survival probability of 98.7(2)\%. Using this method, we rearrange atoms in 3D, implementing a transport sequence where a moving set of atoms leapfrogs over a stationary set. 

\section{Methods}\label{sec:methods}

\subsection{Hologram generation}\label{sec:soeh}

A spatial light modulator (SLM) is a phase-modulating device which imprints a two-dimensional phase distribution $\Phi_\text{SLM}(\bm r')$ on an incident wavefront, where $\bm r'$ is the transverse coordinate in the SLM plane. In a standard optical tweezer setup, the modulated optical field in the SLM plane \begin{equation}
  U_{\rm SLM}(\bm r') = A_{\rm SLM}(\bm r')\,e^{i\Phi_{\rm SLM}(\bm r')}
  \label{eq:uslm}
\end{equation} is optically relayed with magnification $\mathcal{M}$ onto the back aperture of an objective lens of focal length $f$. The field in the focal plane is the Fourier transform of $U_{\rm SLM}$~\cite{goodman1996introduction}, so the focal-plane intensity is\begin{equation}
  I_{\rm F}(\bm r) = \bigl|\mathcal{F}[U_{\rm SLM}](\bm r)\bigr|^2 .
  \label{eq:if}
\end{equation} 
Because the SLM only modulates the phase, $A_\text{SLM}$ is fixed by the intensity profile of the incident laser beam. Determining the $\Phi_\text{SLM}$ that produces a target focal-plane intensity pattern $I_\text{F}$ subject to this constraint is the phase-retrieval problem, which is typically tackled using iterative Fourier transform algorithms \cite{gerchberg1972practical, fienup1982phase, sinclair2004interactive}.

The phase-retrieval problem simplifies when the target is an array of small spots because its inverse Fourier transform can be written in closed form. Consider a single trap of wavelength $\lambda$ and position $\bm r_n=(x_n,y_n)$ in the focal plane, with the origin on the optical axis. The intensity distribution is proportional to $\delta({\bm r}-{\bm r_n})$, and the required field at the SLM is a linear phase gradient $\phi_n(\bm r')=\bm k_n \cdot \bm r'$ described by the wavevector
\begin{equation}
    \bm k_n=\frac{2\pi\mathcal{M}}{\lambda f} \bm r_n.
\label{eq:blazedgrating}
\end{equation}

To extend this to an array of $N$ well-separated traps, we add up the complex fields needed to produce each individual trap, with weights $w_n$ and phase offsets $\theta_n$,
\begin{equation}
  U(\bm r') = \sum_{n=1}^{N} w_n \, e^{i(\bm k_n \cdot \bm r' + \theta_n)}.
  \label{eq:soeh}
\end{equation}
$U(\bm r')$ is the target field in the SLM plane. Since the SLM only modulates the phase, the actual field is Eq.~(\ref{eq:uslm}) with 
\begin{equation}
  \Phi_{\rm SLM}(\bm r') = \arg\bigl[U(\bm r')\bigr].
  \label{eq:trunc}
\end{equation}
The $w_n$ and $\theta_n$ are free parameters that can be optimized to maximize the fraction of light going into the target array, equalize trap depths, and control the optical phases of each trap. This is the SEH method.

To optimize $w_n$ and $\theta_n$, we need to evaluate the field at each trap position in the focal plane. For an SLM with $M$ square pixels, each of side length $a$, located at positions $\bm r'_j$, the field in the focal plane is
\begin{equation}
    U_{\rm F}(\bm r_n) = \mathcal{E}(\bm k_n) \frac{1}{M}\sum_{j=1}^M U_{\rm SLM}(\bm r_j') \, e^{-i\bm k_n\cdot\bm r_j'},
    \label{eq:diffsum}
\end{equation}
where $\mathcal{E}(\bm k_n) = \mathrm{sinc}(k_{nx}a/2)\,\mathrm{sinc}(k_{ny}a/2)$ is the pixel-aperture envelope, which comes from integrating the phase gradient across each pixel and describes the decreasing diffraction efficiency away from the optical axis~\cite{manetsch2025tweezer, chew2024ultraprecise}. We will absorb $\mathcal{E}$ into the weight optimization, so we set it to unity in what follows. The intensity and phase of each trap are \begin{equation}
    I_n=\bigl|U_{\rm F}(\bm r_n)\bigr|^2, \ \ \ \ \ \phi_n = \arg[U_{\rm F}(\bm r_n)].
    \label{eq:evaltrapintensityphase}
\end{equation}
Here, $\phi_n$ denotes the actual optical phase of trap $n$, whilst $\theta_n$ represents the target phase offset.

To implement the algorithm, it is convenient to rewrite Eqns.~(\ref{eq:diffsum}) and (\ref{eq:soeh}) as matrix-vector products,
\begin{align}
    \bm U_{\rm F} &= \frac{1}{M}\,G^T\,\bm U_{\rm SLM}
    \label{eq:propforward} \\
    \bm U &= G^*\bm c
    \label{eq:propbackward}
\end{align}
where we have defined a ($M\,\text{pixels}\,\times\,N\,\text{traps}$) propagation matrix $G_{jn}=e^{-i\bm k_n\cdot\bm r'_j}$ and a complex coefficient vector $c_n=w_ne^{i\theta_n}$, and $\bm U$, $\bm U_{\rm F}$ and $\bm U_{\rm SLM}$ are now vectors. In this way, each hologram computation and each evaluation of the trap fields involves an $M\times N$ matrix multiplication, which can be calculated efficiently on a GPU.

Starting from a uniform weight distribution $w_n=1$ and random phases $\theta_n$ drawn uniformly on $[0,2\pi)$, each iteration of the optimization procedure proceeds as follows: \begin{enumerate}
    \item Build the ideal SLM field $U$ from Eq.~(\ref{eq:propbackward}).
    \item Impose SLM amplitude constraint $A_\text{SLM}=\sqrt{I_\text{SLM}}$.
    \item Evaluate the trap fields $\bm U_{\rm F}$ from Eq.~(\ref{eq:propforward}), and extract $I_n$ and $\phi_n$ using Eq.~(\ref{eq:evaltrapintensityphase}).
    \item Update trap weights and phases according to
    \begin{equation}
        w_n^{i+1}=w_n^i  \left[\frac{\langle I^i \rangle}{I_n^i}\right]^{\alpha}, \ \ \ \ \ \ \ \ \theta_n^{i+1} = \phi_n^i.
    \label{eq:update}
    \end{equation}
\end{enumerate} 
Here $i$ denotes the iteration and $\alpha$ is a damping parameter, typically set to 0.35. The phase update function continuously evolves the trap phases to maximize diffraction into the target spots, while the weight update function \cite{chew2024ultraprecise, di2007computer} converges towards traps of uniform intensity.

\begin{figure*}
\includegraphics{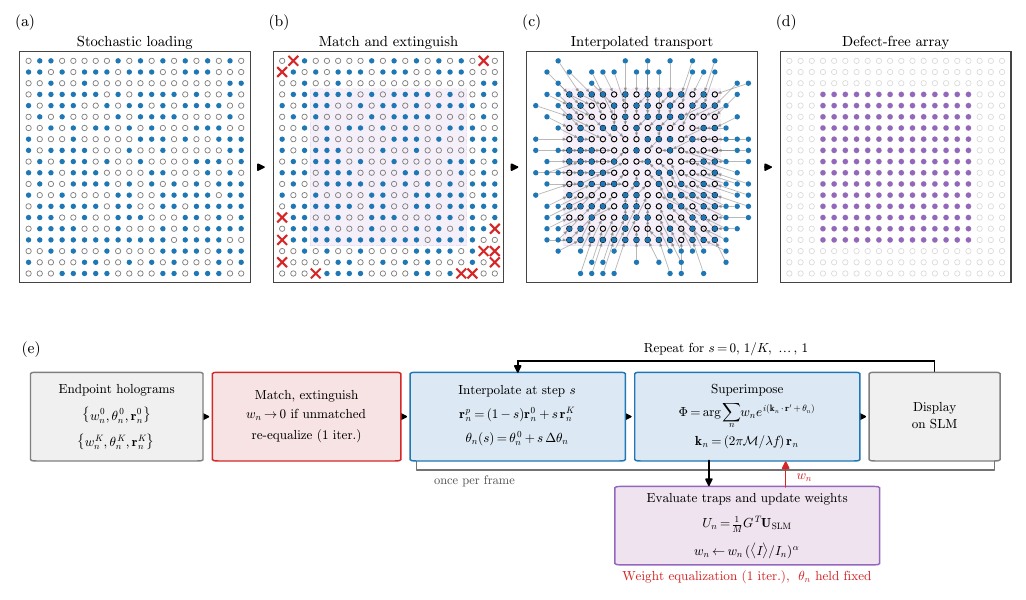}
\caption{\label{fig:wide}Schematic of the I-SEH rearrangement protocol for a stochastically filled $20\times20$ array rearranged into a defect-free $14\times14$ array. (a) Stochastic loading. Filled circles mark occupied sites and open circles empty sites. (b) Occupied sites are assigned to the target sites (shaded region) using the Jonker-Volgenant(JV) linear assignment algorithm. Unassigned and unoccupied traps are extinguished by setting their weights to zero. (c) The assigned traps are transported in parallel along linear trajectories over $K$ frames, with $K$ set by the maximum step size per frame. (d) The final defect-free array. (e) Hologram computation. Weights $w_n$ and phases $\theta_n$ of the initial and final arrays are optimized in advance. After assignment and extinction, the surviving weights are re-equalized with one refinement iteration. For each frame $p = 0, 1, \dots, K$, the trap positions, weights, and phases are linearly interpolated, Eqs. \eqref{eq:interp0} - \eqref{eq:interp2}, and the phase mask $\Phi_{\rm SLM}$ is obtained by Eq.~\eqref{eq:soehlpi}. The field at each trap is evaluated analytically, Eq.~\eqref{eq:propforward}, and the weights are updated with the phases held fixed.}
  \label{fig:protocol}
\end{figure*}

\subsection{Rearrangement}\label{sec:soeh-lpi}

\subsubsection{Linear interpolation}\label{sec:flicker}

Rearrangement requires a \textit{sequence} of holograms. However, when the SLM switches from one hologram to the next, there is a switching transient where the two holograms are displayed simultaneously, and the corresponding optical fields interfere. The transient field in the focal plane has the form 
\begin{equation}
  U(\bm r, t) = a(t)\,A_1(\bm r)\,e^{i\varphi_1}
              + b(t)\,A_2(\bm r)\,e^{i\varphi_2},
  \label{eq:transient}
\end{equation}
where $a(t) + b(t) = 1$ throughout~\cite{Lin2025}. For a trap whose position is unchanged but whose optical phase shifts by $\Delta\varphi$, Eq.~(\ref{eq:transient}) gives $I(t)/I_0=1-4a(t)b(t)\sin^2(\Delta\varphi/2)$, with a minimum at $a=b=1/2$. At this minimum, the intensity is $I_{\min}=I_0\cos^2\!\frac{\Delta\varphi}{2}$, so when $\Delta\varphi=\pi$ the trap is completely extinguished between frames. With a uniformly distributed $\Delta\varphi$, the probability of the relative intensity falling below $x$ is
\begin{equation}
     P\left(\frac{I_{\rm min}}{I_0}<x\right)=\frac{2}{\pi}\sin^{-1}\sqrt{x}.
 \end{equation} 
We find that half the traps will have an intensity dip of 50\% while one in five will fall below 10\%. Since many intermediate holograms are needed for rearrangement, atom loss becomes highly likely.

Now consider a Gaussian beam trap with waist $w_0$ that is displaced by a distance $\delta$ between frames. Provided $\delta$ is sufficiently small such that the trap only has a single minimum as it moves, this minimum still occurs at $a=b=1/2$ when the trap has moved by $\delta/2$, and the intensity is
\begin{equation}
    \frac{I_{\min}}{I_0} = \exp\left(-\frac{\delta^2}{2w_0^2}\right)\cos^2\!\frac{\Delta\varphi}{2}.
    \label{eq:dip2}
\end{equation}
This quantifies the effect of the reduced spatial overlap between successive traps. For a step size equal to the waist, the overlap factor alone gives a normalized transient trap intensity of $60.7\%$. If the step size is too large, the transient trap depth becomes too small, and the atom is lost.

To avoid these losses, we need precise control over the positions and optical phases of every trap. The SEH approach naturally provides the required control since the positions and phases are explicit parameters in Eq.~(\ref{eq:soeh}). We interpolate positions, weights, and phases linearly in $K$ steps between the initial and final holograms. We often refer to the intermediate holograms as ``frames''. For frame $p=0,1...K$, we define $s=p/K$ and write 
\begin{align}
  \bm r_n^p &= (1-s)\,\bm r_n^{0} + s\,\bm r_n^{K},\label{eq:interp0}\\
    \bm k_n^p &= \frac{2\pi\mathcal{M}}{\lambda f}\bm r_n^p,
    \label{eq:interp1}\\
    w_n^{p} &= (1-s) w_n^0 + s w_n^K,\label{eq:weightinterp}\\
  \theta_n^p &= \theta_n^{0} + s\,\Delta\theta_n,\\
    \Delta\theta_n &= \mathrm{wrap}\!\left[\theta_n^{K}-\theta_n^{0}\right]_{-\pi}^{\pi},
    \label{eq:interp2}
\end{align}
where the phase wrapping in Eq.~(\ref{eq:interp2}) ensures the phase interpolation takes the shortest route around the circle between its initial and final values. The hologram at step $p$ follows from Eqs.~(\ref{eq:soeh}) and (\ref{eq:trunc}),
\begin{equation}
  \Phi_{\rm SLM}^{p}(\bm r')
  = \arg\!\left[\sum_{n=1}^{N} w_n^p\,
    e^{i\left(\bm k_n^p\cdot\bm r' + \theta_n^p\right)}\right].
  \label{eq:soehlpi}
\end{equation}
By dividing the change between initial and final phases over $K$ steps, we reduce the per-step phase shift to $\Delta\theta_n/K$, which reduces the transient intensity drop to $I_{\rm min}/I_0=\cos^2(\Delta\theta_n/2K)\approx 1 - (\Delta\theta_n/2K)^2$, falling as $K^{-2}$. In the worst case when $\Delta\theta_n=\pi$, dividing the sequence into $K=10$ steps reduces the intensity drop to only 2.4\%.

Linear interpolation does not optimize the weights of intermediate holograms, resulting in considerable non-uniformity of the trap depths during rearrangement. We can improve on this by implementing a single iteration of the weight update, yielding better weights
\begin{equation}
    (w_n^p)'=w_n^p \left[\frac{\langle I^p\rangle}{I_n^p}\right]^\alpha. \label{eq:weightupdate}
\end{equation}

\subsubsection{The rearrangement protocol}\label{sec:rearrangementprotocol}

Figure \ref{fig:protocol} shows the interpolated superposition of elementary holograms (I-SEH) rearrangement protocol, which proceeds as follows:
\begin{enumerate}
\setlength{\itemsep}{0pt}
    \item Following the procedure described in Sec.~\ref{sec:soeh}, the holograms for the initial and final arrays are computed and optimized prior to rearrangement, yielding
    $\{\bm r_n^{0}, w_n^{0}, \theta_n^{0}\}$ and
    $\{\bm r_n^{K}, w_n^{K}, \theta_n^{K}\}$.
    \item The initial array is stochastically loaded and imaged to extract an occupancy map. [Fig.~\ref{fig:protocol}(a)].
    \item Occupied sites are matched to target sites using the Jonker-Volgenant (JV) linear-sum assignment algorithm that minimizes a global squared-distance cost, suppressing path collisions and bounding the longest trajectory. Unmatched and unoccupied traps are extinguished. [Fig.~\ref{fig:protocol}(b)].
    \item The remaining traps are transported to their target configurations using $K$ frames, computed using the interpolation method described by Eqns.~(\ref{eq:interp0})-(\ref{eq:interp2}) and optionally including the per-step weight optimization described by Eq.~(\ref{eq:weightupdate}). [Fig.~\ref{fig:protocol}(c)].
    \item The final frame in this sequence is the optimized final hologram. [Fig.~\ref{fig:protocol}(d)]
\end{enumerate}  
The rearrangement duration is dominated by step 4.

\subsection{Three-dimensional transport}\label{sec:transport-protocol}

The SEH method is naturally extended to three dimensions with the addition of a trap-dependent quadratic phase term $\frac{\pi\mathcal{M}^2}{\lambda f^2}\,z_n\, |\bm r'|^2 = h_n |\bm r'|^2$, where $z_n$ is the $z$-coordinate of trap $n$ relative to the focal plane. For clarity, we keep $\bm r_n = (x_n,y_n)$ rather than define a 3D position vector. With this additional term Eq.~(\ref{eq:soeh}) becomes
\begin{equation}
  U(\bm r') = \sum_{n=1}^{N} w_n \, e^{i(\bm k_n \cdot \bm r' + h_n|\bm r'|^2+\theta_n)}.
  \label{eq:soeh3D}
\end{equation}
Evaluating the trap fields at their positions follows the same procedure outlined in Sec. \ref{sec:soeh}, with the addition of the inverse quadratic phase term:
\begin{equation}
    U_{\rm F}(\bm r_n, z_n) = \mathcal{E}(\bm k_n) \frac{1}{M}\sum_{j=1}^M U_{\rm SLM}(\bm r_j') \, e^{-i(\bm k_n\cdot\bm r_j' + h_n |\bm r_j'|^2)}.
    \label{eq:diffsum3D}
\end{equation}
Similarly, implementing this algorithmically involves simply modifying the propagation matrix $G_{jn}=e^{-i(\bm k_n\cdot\bm r_j' + h_n|\bm r_j'|^2)}$ to include the quadratic phase term. As such, each evaluation of $\bm U_F$ and computation of $\bm U$ can still be expressed as an $M\times N$ matrix-vector multiplication and remains efficient to compute on a GPU. Optimizing $w_n$ and $\theta_n$ then proceeds in exactly the same way as described in Sec. \ref{sec:soeh}.

For 3D transport, the position of trap $n$ in frame $p$ is $(\bm r_n^p,z_n^p)=\boldsymbol{\gamma}_n(s)$ where $\boldsymbol{\gamma}_n:[0,1]\rightarrow\mathbb{R}^3$ defines the path whose start and end points are $\gamma_n(0)=(\bm r_n^0,z_n^0)$ and $\gamma_n(1)=(\bm r_n^K,z_n^K)$. The hologram at step $p$ is now 
\begin{equation}
  \Phi_{\rm SLM}^{p}(\bm r')
  = \arg\!\left[\sum_{n=1}^{N} w_n^p\,
    e^{i\left(\bm k_n^p\cdot\bm r' + h_n^p|\bm r'|^2 + \theta_n^p\right)}\right],
  \label{eq:soehlpi3D}
\end{equation}
where $h_n^p=\frac{\pi \mathcal{M}^2}{\lambda f^2}\, z_n^p$.

\section{Experimental Results}\label{sec:experimental}

\begin{figure*}[t]
\centering
\includegraphics[width=\textwidth]{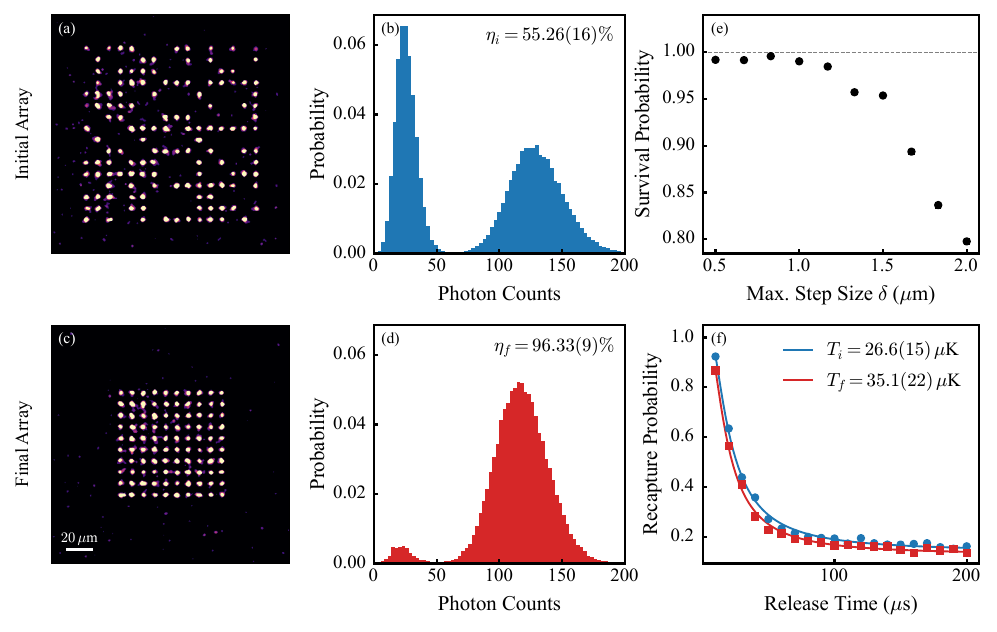}
\caption{Experimental characterization of the rearrangement protocol with zero equalization iterations per step. (a)~Single-shot fluorescence image of a stochastically loaded $16\times16$ array and (b)~the corresponding distribution of photon counts, giving a mean filling fraction $\eta_i = 55.26(16)\%$. (c)~Single-shot fluorescence image of the $10\times10$ target array after rearrangement and (d)~the corresponding count distribution, giving $\eta_f = 96.33(9)\%$. The residual low-count peak in (d) corresponds to target sites left unfilled. (e)~Rearrangement efficiency as a function of the maximum per-frame step size $\delta$. Error bars are smaller than the points. The efficiency is roughly constant for $\delta \lesssim 1.0\,\mu$m and falls for larger steps, as the reduced spatial overlap between successive traps reduces the transient trap depth during the SLM refresh [Eq.~\eqref{eq:dip2}]. (f)~Release-and-recapture measurement of the atomic temperature without (blue circles) and with (red squares) rearrangement (for a $0.5 \,\mu $m max. step size), yielding $T_{\rm i} = 26.6(15)\,\mu$K and $T_{\rm f} = 35.1(22)\,\mu$K.}
\label{fig:experimental}
\end{figure*}

\subsection{Rearrangement in two dimensions}\label{sec:experimental-rearrangement}

We implement the rearrangement protocol described in Sec.~\ref{sec:soeh-lpi} on an array of $^{87}$Rb atoms. Atoms are stochastically loaded from a magneto-optical trap and red-detuned molasses into a $16\times16$ square array of tweezers with $5.0\,\mu$m spacing, generated by the SLM. The tweezers have a wavelength of $\lambda=852$~nm, and a waist of $w_0 = 1.00(5)~\mu$m determined from measurements of the radial trap frequency and the laser power entering the array. The mean trap depth is $U_0/k_B = 480(50)\,\mu$K. We choose a conservative $10\times10$ target array of the same site spacing. At the measured filling fraction, this leaves a wide margin: the $256$ initial sites yield approximately $141 \pm 8$ atoms, so there are always enough atoms to fill the target array.

Site occupancies are determined by fluorescence imaging with a 5\,ms exposure. For each site, we sum the counts in a $9\times9$ pixel region weighted by the measured point-spread function per site, and compare this with a detection threshold set at the crossover between Gaussian fits to the background and single-atom peaks. This yields an imaging fidelity of $99.98\%$ at a baseline imaging survival probability of $99\%$.

The initial and final images are separated by a fixed $50\,$ms interval, within which the rearrangement takes place. Fixing this interval ensures that every experimental run is subject to a common, fixed background loss, so that rearrangement performance can be compared independently of the number of frames used. With no rearrangement, we measure a survival probability of $\eta_0 = 97.01(8)\%$ over this interval, with the loss attributed primarily to collisions with background gas.  

Figure \ref{fig:experimental}(a) shows a single-shot fluorescence image of the initial array, and Fig. \ref{fig:experimental}(c) shows the defect-free array following rearrangement. We do $1000$ repetitions of this rearrangement with a maximum step size of $0.5\,\mu$m and zero intensity equalization iterations per step. The resulting photon count distributions are shown in Figs. \ref{fig:experimental}(b) and \ref{fig:experimental}(d). Applying the computed detection thresholds, we obtain mean filling fractions of $\eta_{\rm i} = 55.26(16)\%$ before and $\eta_{\rm f} = 96.33(9)\%$ after rearrangement. Accounting for the bare survival, the rearrangement survival probability is $\eta = \eta_{\rm f}/\eta_0 = 99.31(12)\%$. This corresponds to a defect-free final array in $50\%$ of iterations immediately following the rearrangement. 

On average, the rearrangement required 20.78 intermediate holograms with a total rearrangement time of 18.23~ms, measured from the moment the initial image is received to the final SLM hologram upload. This includes the time taken to extract an occupancy map of the initial array, perform the JV assignment between occupied and final traps, and then calculate and upload intermediate holograms to the SLM in parallel. 

Figure \ref{fig:experimental}(e) plots the rearrangement survival probability, $\eta$, as a function of the maximum trap displacement between successive frames $\delta$, showing an initial plateau at $\sim99\%$ with a roll-off at $\sim1.0\,\mu\rm m$. The roll-off marks the critical step size at which the transient dip in trap depth, caused by low spatial overlap between successive traps, becomes severe enough for atom loss, as discussed in Sec.~\ref{sec:soeh-lpi}.  

Finally, with $\delta=0.5\,\mu\rm m$, we evaluate whether rearrangement induces any atomic heating. To do this, we perform a release-and-recapture measurement of the atomic temperature with and without rearrangement. In both cases, we measure the temperature after a fixed 50~ms interval following the initial loading. The resulting recapture probability curves for the two cases are shown in Fig.~\ref{fig:experimental}(f). The temperatures before and after rearrangement are $26.6(15)\,\mu\rm K$ and $35.1(22)\,\mu\rm K$, implying an increase of $9(3)\,\mu\rm K$ due to the rearrangement. This increase is small compared to the trap depth.

\begin{figure*}[t]
\centering
\includegraphics[width=\textwidth]{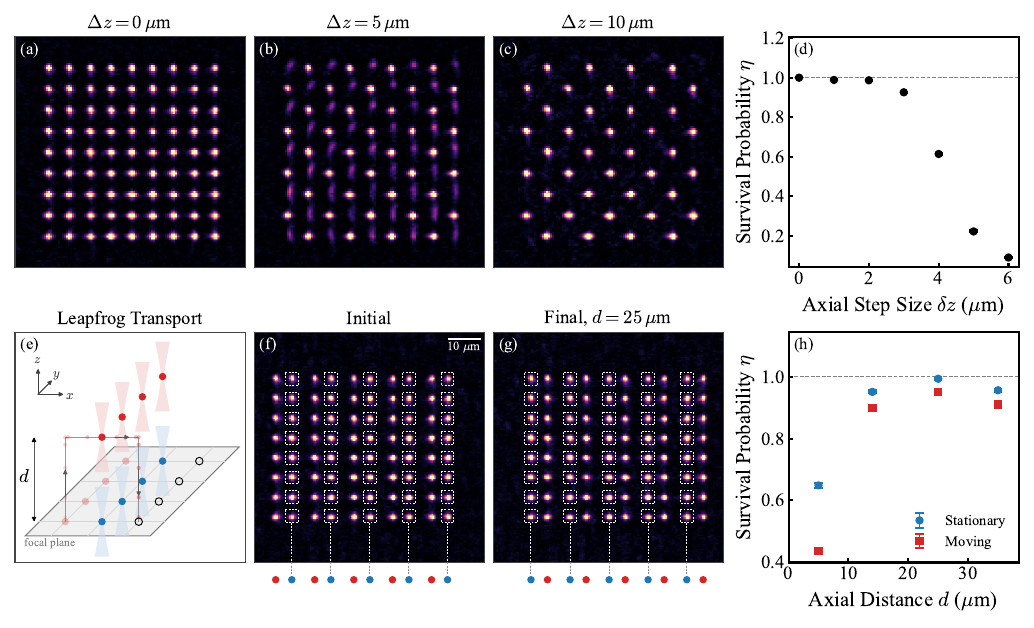}
\caption{Transport out of the focal plane. (a)--(c) Averaged fluorescence images of atoms in a $9\times9$ array with site spacing $5\,\mu$m, with one checkerboard sublattice displaced along the optical axis by $\Delta z = 0$, $5$ and $10\,\mu$m, respectively. The imaging system remains focused on the original focal plane, so atoms in the displaced sites appear defocused. (d) Transport survival probability $\eta$ against axial step size using 20 axial steps of size $\delta z$ out of and back into the focal plane. (e) Schematic of leapfrog transport. Alternate columns (red) are axially transported by $d$, carried over the neighboring column (blue), and back into the focal plane on its far side. (f), (g) Averaged fluorescence images of atoms before (f) and after (g) leapfrog transport with $d = 25\,\mu$m. Dashed squares mark the stationary sites; markers below each image identify the moving (red) and stationary (blue) columns. (h) Survival probability $\eta$ of stationary (blue circles) and moving (red squares) atoms against axial offset plane displacement $d$. In (d) and (h), error bars are smaller than the sizes of the points.}
\label{fig:experimental-transport}
\end{figure*}

\subsection{Transport in three dimensions}

To implement three-dimensional transport using the I-SEH method, we use the same experimental setup as outlined in Sec. \ref{sec:experimental-rearrangement}. For each experimental run, we first assemble a defect-free $9\times9$ tweezer array from an initial stochastically loaded $13\times13$ array, both with a site spacing of 5 $\mu$m, using the rearrangement protocol shown above. We take a 10~ms fluorescence image of this assembled array, then implement a three-dimensional transport sequence following the protocol described in Sec. \ref{sec:transport-protocol}, and finally take a second 10~ms fluorescence image to determine which atoms survive.

First, to demonstrate site-specific transport of tweezers in the axial direction, we move alternating trap sites by $\Delta z$ out of the focal plane in uniform steps of $\delta z=1.0\,\mu$m. Figure \ref{fig:experimental-transport}(a) shows the initial fluorescence image of the array after it has been assembled. Figure \ref{fig:experimental-transport}(b) shows the final fluorescence image of the array after moving alternating traps by $\Delta z=5$~$\mu$m. The transported atoms are still visible as dim spots that are out of focus, while the stationary atoms still appear as sharp, bright spots. Figure \ref{fig:experimental-transport}(c) shows the final image when $\Delta z=10$~$\mu$m. Here, the transported atoms are far enough out of the focal plane that they are no longer visible in the image, but they have not been lost, as we show next.  

We determine the maximum axial step size that can be used before significant atom loss occurs using the following experiment. For a uniform step size $\delta z$, we transport the array by $\Delta z=10\,\delta z$ in the axial direction, hold it there for 10~ms, then transport it back by $\Delta z=-10\,\delta z$ into the focal plane, and then acquire an image. There is no trap intensity equalization applied in this protocol. Figure \ref{fig:experimental-transport}(d) plots the resulting transport survival probability as a function of $\delta z$, showing a roll-off at $\sim 3.0\,\mu$m with a survival probability of 98.7(2)\% for a step size of $\delta z=2.0\,\mu$m. With a trap waist of $w_0=1.00\,\mu$m, we have a Rayleigh range of $z_R=3.69\,\mu$m, assuming an ideal Gaussian beam, which roughly matches the observed roll-off point. 

Three-dimensional transport is most useful if traps can be moved in one axial plane without perturbing atoms held in another. To examine this, we perform a \textit{leapfrog} transport sequence, in which a trap is axially transported by $d$, then moved \textit{behind} a neighbouring trap, and returned to the focal plane on its far side. A diagram of this transport sequence is shown in Fig.~\ref{fig:experimental-transport}(e), with stationary traps in blue and moving traps in red. We test this experimentally using an assembled 80-site array arranged into five pairs of columns with a $5.0\,\mu$m separation. For each pair, the leftmost column is transported axially by $d$, moved behind its neighbouring column, and then returned to the focal plane on its far side. Figure \ref{fig:experimental-transport}(f) shows an averaged fluorescence image of the initial assembled array, while Fig.~\ref{fig:experimental-transport}(g) shows an averaged image after a leapfrog transport with $d=25\,\mu$m. The stationary traps are marked by dashed squares, and every column of traps is aligned with a blue or red circle beneath each image to indicate if it is stationary or moving. 

For varying axial distances $d$, we measure the survival probability of the stationary and moving traps separately, following the leapfrog transport sequence. We use an axial step size of $\delta z=1.0\,\mu$m and a radial step size of $\delta=0.5\,\mu$m.
We also perform a single iteration of intensity equalization per transport step. Figure \ref {fig:experimental-transport}(h) shows the survival probabilities of the stationary and moving traps as a function of axial distance $d$. We observe a sharp increase in survival probability above $d=10.0\,\mu$m, with a maximum of 99.4(3)\% and 95.1(4)\% for the stationary and moving traps, respectively. Across this range of axial distances, the survival probabilities of the stationary and moving traps are highly correlated. We attribute the low survival probabilities observed at $d=5.0\,\mu$m to two causes. The first is the severe perturbation or optical cross-talk between the moving and stationary traps due to their close proximity. The second is that the traps effectively merge so that two atoms can end up in the same trap and then be rapidly ejected by light-assisted collisions when the imaging light is turned on. We also observe a correlated decrease in survival probability at $d=35.0\,\mu$m that cannot be due to optical cross-talk. It could be due to aberrations of the traps when they are moved so far out of the plane, or it may reflect the greater number of intermediate holograms in this 71-frame transport sequence. 

Finally, we compare the leapfrog transport sequence with and without a single iteration of intensity equalization per transport step. We find that without intensity equalization the maximum survival probability is 10-20\% smaller, illustrating the value of this step of the algorithm.

\begin{figure*}
\centering
\includegraphics[width=\textwidth]{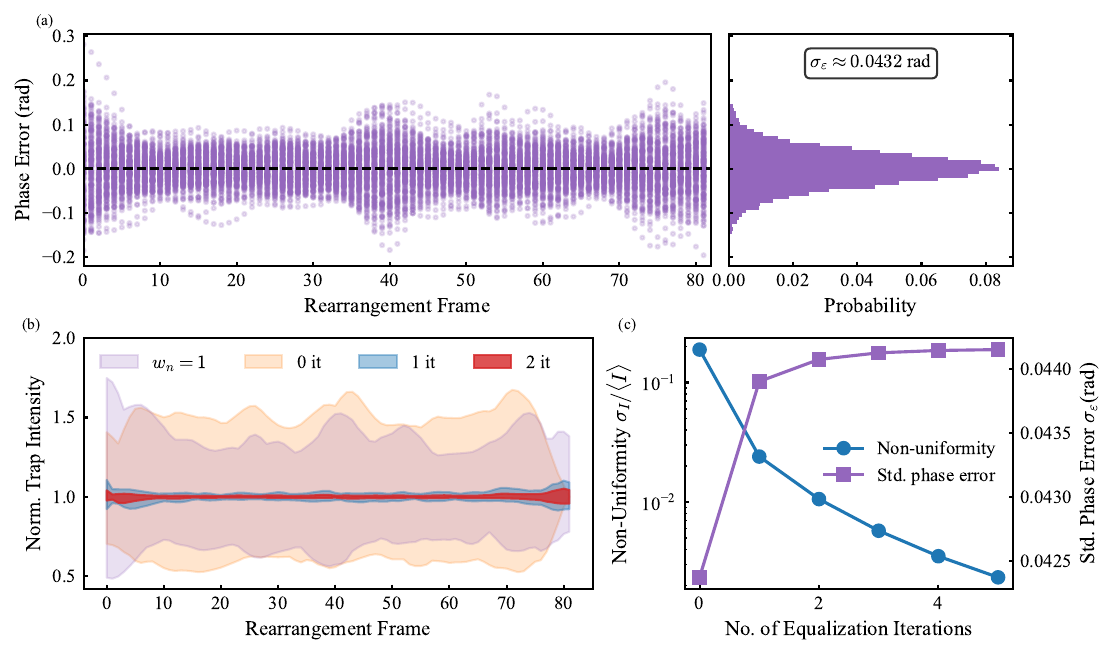}
\caption{\label{fig:wide}Numerical characterization of the I-SEH protocol, simulated for the rearrangement of a $50\%$ stochastically filled $20\times20$ array into a defect-free $14\times14$ array, both of $5.0\,\mu$m site spacing, with a maximum step size of $0.25\,\mu$m and $K=85$ frames. (a)~Phase error $\varepsilon_n$, the deviation of each trap's optical phase from its linearly interpolated target trajectory, shown for all $196$ traps at every frame (left) and as the distribution of all measurements (right). The dashed line marks zero error. The standard deviation over the full sequence is $\sigma_\varepsilon = 0.0432\,\mathrm{rad}$. (b)~Minimum-to-maximum envelope of the normalized trap intensities $I_n/\langle I\rangle$ at each frame, for four weight schemes: fixed uniform weights, and linearly interpolated weights refined by $0$, $1$ and $2$ equalization iterations per frame. (c)~Trap-intensity non-uniformity (blue) $\sigma_I/\langle I\rangle$ and standard deviation in phase-error (purple) against the number of equalization iterations per frame. }
\label{fig:numerics}
\end{figure*}

\section{Numerical Characterisation}\label{sec:numerical}

The protocol outlined in Sec. \ref{sec:soeh-lpi} controls both the trap intensities and their optical phases in each frame of a rearrangement sequence. 
Since these are difficult to measure directly in the experiment, we characterize the performance of the holograms generated using this protocol by numerical simulation. 

We simulate the rearrangement of a 50\% stochastically filled $20\times20$ array with a $5.0 \,\mu\rm m$ site spacing into a defect-free $14\times14$ array with the same site spacing, and set a maximum step size of $0.25 \, \mu\rm m$. We use the parameters of the real experiment: a $1024\times1024$ SLM with a 17 $\mu\rm m$ pixel pitch and a side length of $L=17.41$~mm, a trapping wavelength $\lambda=852 \, \rm nm$, and an objective with a numerical aperture of $\rm NA=0.55$ and effective focal length of $f=17.3$~mm, taking into account the magnification of our optical relay.

\subsection{Phase Control}

Over an $85$ step sequence, we evaluate the optical phase of each trap in the array and calculate the deviation of the trap phases from their linearly interpolated target values, which we call the phase error $\varepsilon_n$. Figure~\ref{fig:numerics}(a) shows the phase error of every trap in each rearrangement frame, with the distribution of all measurements shown on the right. The standard deviation of the phase error is $\sigma_\varepsilon = 0.0432\,\rm rad$, which compares favorably with the $0.2\,\rm rad$ reported for the CNN-based protocol of Ref.~\cite{Lin2025}. With $K=85$, the mean phase shift per step is $0.021\,\rm rad$ and the maximum is $0.079\,\rm rad$. The drop in intensity due to these phase changes is negligible, and is instead dominated by the degree of spatial overlap. For the step size simulated here, this corresponds to $I_{\rm min}/I_0 = 0.879$.

\subsection{Intensity Control}\label{sec:intensitycontrol}

For the same 85 step rearrangement sequence, we evaluate the normalized trap intensities $I_n^p/\langle I^p\rangle$ at each frame and compute the root-mean-square (RMS) non-uniformity $\sigma_I/\langle I\rangle$. We compare four weight update schemes: fixed uniform weights $w_n^p = 1$, and linearly interpolated weights with $0$, $1$ and $2$ equalization iterations per frame, following Eqs.~\eqref{eq:weightinterp} and~\eqref{eq:weightupdate}. The results are summarised in Fig. \ref{fig:numerics}(b)--(c).

Figure \ref{fig:numerics}(b) shows the minimum-to-maximum intensity envelope at each frame of the rearrangement, for the four different weight update schemes. We find that interpolating the weights with no subsequent equalization iterations yields an RMS non-uniformity of $19.85\%$ and a minimum trap intensity of $35.05\%$ of the mean, significantly worse than simply using fixed uniform weights where the non-uniformity is $13.45\%$ and the minimum intensity is $39.88\%$. However, a single equalization iteration reduces the non-uniformity by an order of magnitude to just $1.24\%$, with a minimum trap intensity of $90.55\%$. The effect of further equalization iterations on trap non-uniformity is shown in Fig. \ref{fig:numerics}(c). Subsequent iterations continue to reduce the non-uniformity, but by diminishing amounts. After two iterations, we obtain a non-uniformity of $0.48\%$, which drops to $0.26\%$ after five iterations. 

We also assess whether there is a trade-off between intensity and phase control during rearrangement by examining the degree of phase control obtained in each of the four weight update schemes. We find that the standard deviation of the phase error is $\sigma_{\epsilon}=0.0434$~rad for both fixed weights and for linear interpolation without equalization.  Figure \ref{fig:numerics}(c) shows that $\sigma_{\epsilon}$ increases only marginally when one or more intensity equalization iterations are implemented. Similarly, we find a negligible difference in both the mean and maximum phase shift per step across the four schemes. We see that the method is highly effective at equalizing trap intensities with negligible detriment to the phase control.

\begin{figure*}[t]
\centering
\includegraphics[width=\textwidth]{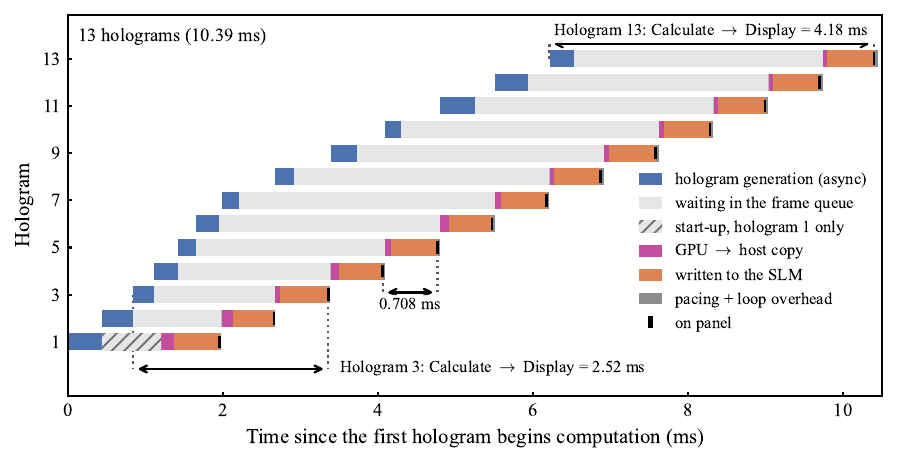}
\caption{\label{fig:chronogram}Timeline of one rearrangement with 13 intermediate
holograms ($10\times10\rightarrow6\times6$ array, $\delta = 0.8\,\mu$m). Each row is
one hologram, divided into the stages of Table~\ref{tab:budget} together with the
time spent waiting in the queue; ticks mark the completion of each write.
Holograms are delivered every $0.708\,$ms (arrow), a rate set by the write and
pacing hold rather than by computation. As computation is faster than delivery,
holograms wait progressively longer in the queue until it is full, raising the
time from computation to display to $4.18\,$ms without changing the delivery rate (queue depth = 5). The hatched segment of hologram 1 is a one-off start-up cost of $0.77\,$ms, part
of the fixed cost discussed in Sec.~\ref{sec:scaling_k}, rather than time spent
queueing. Time zero is the start of computation of the first hologram, $2.61\,$ms after the
initial image arrives.}
\end{figure*}

\section{Protocol Timing Breakdown}\label{sec:timings}

The rearrangement speed is set by whichever is slower, the computation of an intermediate hologram or the time it takes to be displayed on the SLM. The SLM has a minimum refresh period of 0.696 ms, which determines the shortest interval successive holograms can be displayed. In this section we resolve the per-frame cost of each stage of the protocol and determine the array size at which hologram computation with the SEH approach, rather than the SLM, is limiting.

Holograms are computed on a single NVIDIA RTX 5090 (32 GB GDDR7, PCIe Gen5
$\times16$) hosted by a 32-core AMD Threadripper PRO 9975WX. The SLM controller (Meadowlark Optics
UHSPDM1K-700-900-PC8)  board is connected over PCIe Gen3 $\times8$ and holds two image
buffers, so one hologram may be displayed while the next is written into the
other buffer. Host intervals are timed spans
between calls to a high-resolution performance counter of $0.1\,\mu$s resolution;
GPU work is timed with CUDA events recorded around the computation and read only
after the sequence completes, so that reading timings cannot disturb the task being
measured.
Unless stated otherwise, all results in this section are medians over holograms drawn from rearrangements of the
$16\times16 \rightarrow 10\times10$ geometry of Sec.~\ref{sec:experimental}. The
first hologram of each sequence is excluded throughout: it is computed with cold
GPU kernels, copied into a buffer that has just been allocated and written to an SLM that has not yet been addressed, so costs more at every stage of the protocol than the
frames that follow. These one-off costs are treated separately in
Sec.~\ref{sec:scaling_k}. The remaining frames set the delivery rate, and we refer to then as `steady state' holograms. We define the rearrangement time as the interval from
the initial fluorescence image arriving in host memory to the final hologram write
completing, so the camera exposure and readout lie outside the 
figures quoted here.

\subsection{Pipelined Computation and Delivery}

Each intermediate hologram passes through four stages. The interpolated trap
parameters of Eqs.~\eqref{eq:interp0}-\eqref{eq:interp2} are advanced one
step; the SEH sum of Eq.~\eqref{eq:soehlpi} is evaluated and its argument taken;
a fixed, pre-built hologram used for focus adjustment and aberration correction, common to every frame,
is added modulo $2\pi$; and the result is cast to the 8-bit
format the SLM controller board accepts. The finished hologram is then copied from the GPU
into host memory and written to the SLM.

We split this work across two threads. A generator thread advances the trap parameters, evaluates the hologram on the GPU and hands each finished frame off into a queue before beginning the next. A writer thread takes the frames out of the queue in order, copying each into host memory, and issues a write command to the SLM. The copy out of the queue from GPU $\to$ host memory lands into a page-locked host buffer, allocated once per rearrangement in $0.047\,$ms and reused for every subsequent frame, removing any overhead associated with pageable memory management.  The same copy of the 1.05~MB hologram into ordinary pageable memory takes $0.152 \pm
0.004\,$ms, against $0.056\, \pm 0.002\,$ms page-locked. The queue between the two threads is bounded at five holograms. The bound does not
limit the delivery rate as the writer waits only $3\,\mu$s per frame for a
computed hologram. Pipelining the operations in this way, together with the write pacing described below, brings the interval between displayed holograms to almost exactly the SLM refresh rate.

\subsection{Write Pacing}

The controller board acknowledges a hologram write command once the data reaches board memory, which occurs faster than the liquid crystal refresh rate of the SLM. This means that the writer thread can begin over-writing a previous frame before it has been fully realised. Atoms would then skip intermediate positions and move in steps larger than $\delta$, without reported error. The writer thread is therefore paced: each write command is held until $0.700\,$ms have elapsed since the previous one, the $0.696\,$ms SLM refresh plus $4\,\mu$s of margin. Of all the timed rearrangements studied here, no holograms were delivered faster than the paced $0.700\,$ms floor, with the fastest observed being $0.7021\,$ms. 

\subsection{Per Frame Timing}

\begin{table}[tb]
\caption{\label{tab:budget}Per-frame cost of hologram generation and delivery.
The computation of each hologram runs in parallel with the delivery of the
previous one and does not add to the interval between displayed holograms; the
remaining stages run in series on the writer thread. Values are means over 5221
steady-state holograms from 493 rearrangements, with the first hologram of each
excluded.}
\begin{ruledtabular}
\begin{tabular}{lc}
Stage & Time (ms) \\
\colrule
Hologram computation (in parallel) & 0.284 \\
\colrule
GPU synchronization & 0.003 \\
Transfer from GPU to host memory & 0.068 \\
Write to SLM & 0.587 \\
Pacing hold & 0.037 \\
Remaining overhead & 0.013 \\
\colrule
Interval between displayed holograms & 0.708 \\
\end{tabular}
\end{ruledtabular}
\end{table}

Figure \ref{fig:chronogram} shows an example of the timeline for delivering 13 intermediate holograms. Since hologram computation is faster than its delivery, each hologram waits
about $0.4\,$ms longer in the queue than the last, until the queue fills. The timing diagram illustrates how the rearrangement speed is limited by the SLM rather than by computation. The delay between computation and display is capped at about
$4.2\,$ms. Table~\ref{tab:budget} shows the delivery interval broken into its stages, averaged over 493 rearrangement cycles. Writing to the SLM takes up the majority of the interval at $0.587\,$ms. Without intensity equalisation, hologram computation is
$0.284\,$ms and the copy from GPU $\to$ host is $0.068\,$ms, both of which are pipelined and run concurrently with the write, so do not add to the total interval. The GPU synchronization is the time spent waiting for a computed hologram, and the
remaining overhead is the untimed handover between the two threads.

\subsection{Scaling with Sequence Length}\label{sec:scaling_k}
Fitting the rearrangement time against the number of
intermediate holograms $K$, spanning $K = 4$ to $23$, gives a cost of $0.697 \pm 0.003\,$ms per hologram without equalization and a fixed cost of $3.741 \pm 0.039\,$ms, with
$R^2 = 0.990$. The slope lies within $1\,\mu$s of the $0.696\,$ms refresh period,
so the cost of additional frames is set by the SLM. The delivered
interval is stable to a robust standard deviation of $2.2\,\mu$s between
rearrangements, with some longer intervals when the operating system
interrupts the writer.

The fixed cost is incurred once per rearrangement and is mainly the time taken by the
assignment and interpolation setup. On average, the first hologram reaches the SLM $4.32\,$ms after the initial image arrives; extrapolating the fit to $K = 0$ removes one delivered interval, giving $3.61\,$ms, consistent with the fitted value.

\subsection{Scaling with array size}\label{sec:scaling}

If the longest trajectory in a rearrangement is $d_{\max}$, the number of holograms required is $K = \lceil d_{\max}/\delta \rceil$, and does not depend on how the holograms are computed. For a target array occupying a fixed fraction of the initial array, at a fixed site spacing and filling fraction, $d_{\max}$ grows roughly with the linear size of the array, so $K \propto \sqrt{N}$. For the $16\times16 \rightarrow 10\times10$ rearrangement of Sec.~\ref{sec:experimental}, $K \approx 21$ at $\delta = 0.5\,\mu$m, and doubling the width of both arrays would roughly double it. Provided each hologram is computed within the refresh period of the SLM, the time taken to move the atoms therefore grows as $\sqrt{N}$, which is a better scaling than the linear growth of sequential rearrangement methods. 

Each SEH hologram is computed from the $M\times N$ matrix products of
Eqs.~\eqref{eq:propforward} and~\eqref{eq:propbackward}, so the cost scales as $\mathcal{O}(MN)$  whereas an FFT-based hologram costs $\mathcal{O}(M\log M)$
independent of the number of traps~\cite{Cooley1965}. This means that for a large enough array size the SEH approach will become the slower of the two. Here we determine where this crossover in times lies, how this changes with the required positional accuracy, and the array size where hologram computation remains below the SLM refresh period.  We
time the full per-frame computation of Sec.~\ref{sec:timings} on the same GPU,
averaged over five repetitions, for array sizes ranging from $N = 4$ to $3\times10^5$.

\begin{figure}
\includegraphics[width=\columnwidth]{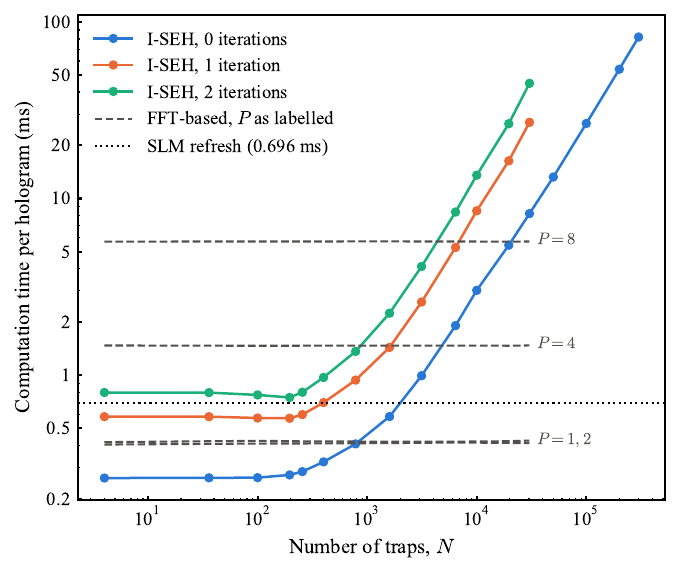}
\caption{\label{fig:scaling}Computation time per hologram against the number of 
traps $N$ in the final array. SEH holograms with zero, one and two weight-equalization iterations
(solid), and FFT-based holograms at zero-padding factors $P = 1$, $2$, $4$ and $8$
(dashed), whose cost is independent of $N$. The dotted horizontal line marks the
$0.696\,$ms SLM refresh period. Means over five repeats.}
\end{figure}

Figure \ref{fig:scaling} shows the results of this analysis. Below $N \approx 200$, the computation time is set by a fixed overhead
paid on every hologram for dispatching work to the GPU, rather than by the number
of traps. For $N \gtrsim 1000$, the time grows linearly, at about $0.27\,\mu$s per trap. With no equalization
iterations, computation remains within the $0.696\,$ms refresh period
for arrays up to $N \approx 2100$, and with one iteration up to
$N \approx 435$. Two iterations
exceed the refresh period at every array size, as their fixed cost is
$0.70$--$0.74\,$ms, and take $0.78\,$ms at $N = 256$. 

For comparison to the SEH method, we time a full field FFT-based LPI implementation, in which the intermediate holograms are obtained from a single inverse FFT on the interpolated focal plane field, all done on the same GPU. Since the entire focal plane is transformed, it is independent of the trap number $N$ but grows with a zero-padding factor $P$, which is required for finer trap site placement. At $P=1$ or 2, the FFT implementation is faster than SEH for $N \gtrsim 800$. In our setup, the Fourier unit is $\Delta r = \lambda f/L = 0.847 \ \mu{\rm m}$. Thus, without zero padding ($P=1$), the \textit{minimum} step size along a diagonal, dictated by the spatial resolution of the FFT, is 1.2~$\mu$m. This is larger than the trap waist, and beyond the \textit{maximum} step size where we start to observe atom loss, as shown in Fig.~\ref{fig:experimental}(e). Thus, some zero padding is necessary in an FFT-based implementation. At $P=4$ the minimum step size along a diagonal is 0.3~$\mu$m, similar to the maximum step sizes we have used. In this case, the FFT approach is faster than SEH for $N \gtrsim 5000$.

\section{Conclusions}\label{sec:conclusion}

We have demonstrated real-time parallel rearrangement of neutral atoms in both two and three dimensions, using a high-speed SLM and an algorithm that combines linear phase interpolation with the superposition-of-elementary-holograms method. We call this the I-SEH algorithm. The transverse position of each trap is specified in the hologram through a phase term that varies linearly in the SLM coordinate, and its axial position through a phase term that varies quadratically. These are continuous parameters so tweezer positions are not fixed to a computational grid, and sub-Fourier-unit trap positioning and transport steps can be done without zero-padding. Higher order terms could also be added, e.g. for per-site aberration correction. In simulation, the protocol controls the optical phase of every tweezer to within $\sigma_\varepsilon = 0.045\,$rad of its target linear trajectory, while two weight-equalization iterations per rearrangement step reduce the trap-intensity non-uniformity of intermediate holograms from ${\sim}20\%$ to $0.48\%$, with a negligible cost in phase control. As such, the transient trap-intensity dip during each SLM refresh cycle is dominated by the spatial overlap of successive trapping potentials rather than by their phase difference. 

In steady state, and without intensity equalization, each hologram takes an average of $0.284$~ms to compute on a single GPU, well within the SLM refresh cycle, so that for arrays of up to ${\sim}2200$ traps the total rearrangement rate is hardware limited rather than constrained by hologram computation. Experimentally, we assemble $100$-atom arrays from stochastically loaded $256$-site arrays in $18.2\,$ms with a rearrangement survival probability of $99.31(12)\%$ per atom and an atomic temperature increase of $9\,\mu$K, more than an order of magnitude below the $480\,\mu$K trap depth, so atoms remain cold and deeply confined after rearrangement. We have shown that a subset of the atoms can be moved axially. When they are moved out of the plane by 20~$\mu$m in steps of 2~$\mu$m, and then back again, the survival probability is 98.7(2)\%. Using both axial and transverse motion, we have demonstrated a 3D rearrangement protocol where moving atoms leapfrog over stationary ones. Here, we measure a survival probability of 99.4(3)\% for the stationary atoms and 95.1(4)\% for the moving ones.

Several interesting directions follow from this work. Of particular interest is whether qubit coherence is preserved during transport. In alkali atoms, the qubit states are usually the two $m_F=0$ components of the ground-state hyperfine interval. These states have a small differential light shift, so a spread of intensities leads to qubit dephasing. In a trap of depth $U_0$ the decoherence rate due to the temperature $T$ is proportional to $k_B T/U_0$. Intensity non-uniformity across the traps is a second source of decoherence, proportional to the fractional spread of intensities $\sigma_{I}/\langle I \rangle$ with the same proportionality constant. Thus, we expect that transport will not substantially change the decoherence rate when $\sigma_{I}/\langle I \rangle < k_{B}T/U_0$. This latter ratio is typically about 5\%, whereas rearrangement with a single intensity equalization at each step yields non-uniformity close to 1\%. Thus, we can expect the coherence time during transport to be similar to the coherence time in static traps.

The high positional accuracy and 3D motion offered by our protocol lends itself to more complex trajectories. This can be helpful for avoiding loss arising from collisions between particles or from interference between traps that get too close. 
Collision avoidance can be incorporated into the JV linear assignment algorithm as a minimum-separation constraint and realized through smooth detours, which in an FFT-based implementation would be quantized to the computational grid. It will be interesting to explore how many different axial planes can be used during a rearrangement sequence and to investigate which 3D transport protocols have the highest fidelity.

The hologram computation has room to go faster. Extinguished traps currently remain in the sum with zero weight and could simply be removed. Below ${\sim}200$ traps the cost is set by the fixed overhead of launching GPU kernels, which could be reduced using fusing operations or by capturing each frame as a single CUDA graph. Since holograms without intensity equalization are independent of one another, frames could be distributed across several GPUs, extending the array size where rearrangement time remains limited by the SLM. Our work motivates the development of faster SLMs. The transport method remains adiabatic until the SLM refresh rate approaches the trap frequencies which are all more than 10 times higher than the refresh rate of the SLM used here.

These methods could be used in continuous loading architectures where atoms in a storage zone are continuously replenished from a reservoir~\cite{Chiu2025}. They can also be used to assemble two- or three-dimensional arrays of molecules~\cite{Anderegg2019, Holland2023, Bao2023} for applications in quantum computing and simulation~\cite{Cornish2024}. The stochastic filling fraction for molecules is lower than for atoms, typically 30-35\%, so there is a greater rearrangement overhead and rearranging in parallel is especially beneficial. The dipolar interaction between molecules is much weaker than between Rydberg atoms, so it is important to bring the molecules as close as possible. It will be interesting to study how small a site spacing can be produced with the methods used here. Finally, the methods seem suitable for transporting auxiliary qubits through an array of stationary data qubits, for example in dual-species arrays of atoms~\cite{Anand2024} or of atoms and molecules~\cite{Zhang2022b, Wang2022, Guttridge2023, Ruttley2026arxiv}. This is especially important for implementing error correction and non-destructive readout protocols.

\begin{acknowledgments}
We are grateful to Thomas Walker, Chi Zhang and Ben Sauer from the Centre for Cold Matter and to Tom Hepworth, Alex Guttridge, Simon Cornish and the whole Cornish research group for helpful advice and discussions that have supported this work. The research has received funding from the UK Engineering and Physical Sciences Research Council (EPSRC) through grants EP/P01058X/1, EP/W00299X/1 and UKRI2226.
KKV gratefully acknowledges the Alexander von Humboldt Foundation for a Feodor Lynen Research Fellowship.
\end{acknowledgments}

\bibliography{apssamp, references}

\end{document}